\documentclass[letterpaper,english,american,prl,twocolumn,amsmath,amssymb,showpacs,superscriptaddress]{revtex4-1}
\usepackage[T1]{fontenc}
\usepackage{babel}
\usepackage{prettyref}
\usepackage{amsbsy}
\usepackage{graphicx}
\usepackage[unicode=true,
 bookmarks=true,bookmarksnumbered=false,bookmarksopen=false,
 breaklinks=false,pdfborder={0 0 0},backref=false,colorlinks=false]
 {hyperref}
\hypersetup{pdftitle={Distribution of pair-wise covariances in neuronal networks},
 pdfauthor={Dahmen, Helias}}
\usepackage{breakurl}

\makeatletter

\newrefformat{cap}{\hyperref[#1]{Figure~\ref{#1}}}
\newrefformat{fig}{\hyperref[#1]{Fig.~\ref{#1}}}
\newrefformat{tab}{\hyperref[#1]{Table ~\ref{#1}}}
\newrefformat{sec}{\hyperref[#1]{Section~\ref{#1}}}
\newrefformat{sub}{\hyperref[#1]{Section~\ref{#1}}}
\newrefformat{cha}{\hyperref[#1]{Chapter~\ref{#1}}}
\newrefformat{eq}{\hyperref[#1]{Eq.~(\ref{#1})}}

\makeatother

\begin{document}
\global\long\def\ms{\:\mathrm{ms}}
\global\long\def\M{\mathbf{M}}
\global\long\def\I{\mathbf{I}}
\global\long\def\Var{\mathrm{Var}}
\global\long\def\E{\mathrm{E}}
\global\long\def\D{\mathbf{D}}
\global\long\def\a{\mathbf{a}}
\global\long\def\w{\mathbf{w}}
\global\long\def\v{\mathbf{v}}
\global\long\def\erfc{\mathrm{erfc}}
\global\long\def\n{\mathbf{n}}
\global\long\def\q{\mathbf{q}}
\global\long\def\diag{\mathrm{diag}}
\global\long\def\x{\mathbf{x}}
\global\long\def\t{\mathbf{t}}
\global\long\def\tr{\mathrm{tr}}
\global\long\def\mD{\mathcal{D}}

\title{Distributions of covariances as a window into the operational regime
of neuronal networks}

\author{David Dahmen}

\affiliation{Institute of Neuroscience and Medicine (INM-6) and Institute for
Advanced Simulation (IAS-6) and JARA BRAIN Institute I, Jülich Research
Centre, Jülich, Germany}

\author{Markus Diesmann}

\affiliation{Institute of Neuroscience and Medicine (INM-6) and Institute for
Advanced Simulation (IAS-6) and JARA BRAIN Institute I, Jülich Research
Centre, Jülich, Germany}

\affiliation{Department of Psychiatry, Psychotherapy and Psychosomatics, Medical
Faculty, RWTH Aachen University, Aachen, Germany }

\affiliation{Department of Physics, Faculty 1, RWTH Aachen University, Aachen,
Germany}

\author{Moritz Helias}

\affiliation{Institute of Neuroscience and Medicine (INM-6) and Institute for
Advanced Simulation (IAS-6) and JARA BRAIN Institute I, Jülich Research
Centre, Jülich, Germany}

\affiliation{Department of Physics, Faculty 1, RWTH Aachen University, Aachen,
Germany}

\date{\today}

\pacs{87.19.lj, 64.60.an, 75.10.Nr, 05.40.-a }
\begin{abstract}

Massively parallel recordings of spiking activity in cortical networks
show that covariances vary widely across pairs of neurons. Their low
average is well understood, but an explanation for the wide distribution
in relation to the static (quenched) disorder of the connectivity
in recurrent random networks was so far elusive. We here derive a
finite-size mean-field theory that reduces a disordered to a highly
symmetric network with fluctuating auxiliary fields. The exposed analytical
relation between the statistics of connections and the statistics
of pairwise covariances shows that both, average and dispersion of
the latter, diverge at a critical coupling. At this point, a network
of nonlinear units transits from regular to chaotic dynamics. Applying
these results to recordings from the mammalian brain suggests its
operation close to this edge of criticality. 
\end{abstract}
\maketitle

\paragraph{}

\begin{figure}
\begin{centering}
\includegraphics{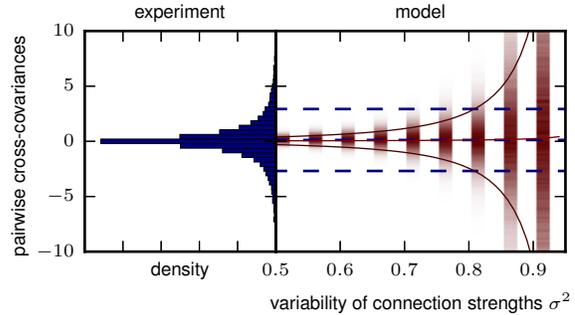}
\par\end{centering}

\centering{}\caption{Distribution of cross-covariances $c_{ij}=\frac{1}{T}\left(\langle n_{i}n_{j}\rangle-\langle n_{i}\rangle\langle n_{j}\rangle\right)$
between spike counts $n_{i}$ in macaque motor cortex (blue histogram)
as compared to integral cross-covariances in homogeneous networks
of Ornstein-Uhlenbeck processes with different variability $\sigma^{2}$
of connection strengths (red shading indicates density of histogram,
red curves the analytical prediction for mean and $\pm1$ standard
deviation). The low mean and large standard deviation (blue dashed
horizontal lines) of experimentally observed cross-covariances (blue)
are explained by a model network (red) with high variability of connections
($\sigma^{2}\approx0.8$). The experimental data from $155$ neurons
are recorded with a $100$-electrode ``Utah'' array (Blackrock Microsystems,
Salt Lake City, UT, USA) with $400\,\mathrm{\mu m}$ interelectrode
distance, covering an area of $4\times4$ mm$^{2}$. Spike counts
$n_{i}$ of activity are obtained within $T=400\protect\ms$ after
trial start (TS) of a reach-to-grasp task \citep{Riehle13_48} in
$141$ trials (subsession: i140703-001). The data set is also available
with all details on the recording and annotations in Brochier et al.
(to be submitted to Scientific Data). Numerical solution of \prettyref{eq:covs}
and analytical predictions for the mean \eqref{eq:mean_cij} and standard
deviation \eqref{eq:var_cij} are performed with network size $N=1000$,
uniform noise $D=2.97$, Gaussian connectivity $\mathcal{N}(\mu=6.5\cdot10^{-4},\sigma^{2}/N)$,
and varying $\sigma^{2}$ (right horizontal axis). Data courtesy of
A. Riehle and T. Brochier. \label{fig:distributions}}
\end{figure}

A network of neurons constitutes a many-particle system with interactions
mediated by directed (asymmetric) and random synaptic connections.
This quenched randomness is the defining feature of a disordered system.
The large number and divergence of outgoing connections implies that
each pair of neurons receives a substantial amount of common inputs,
leading to positively correlated activity on average \citep{DeLaRocha07_802,Shea-Brown08}.
The dominant negative feedback, which stabilizes the strongly fluctuating
activity in recurrent networks \citep{Vreeswijk96}, gives rise to
active decorrelation \citep{Tetzlaff12_e1002596}. As a consequence,
correlations are positive on average, but close to zero \citep{Ecker10,Cohen11_811}
and their mean is predicted to vanish in inverse proportion to the
number of neurons in the network \citep{Renart10_587}. On the level
of individual pairs of neurons, however, experimentally a wide distribution
of correlations is observed, shown in \prettyref{fig:distributions}
for recordings in macaque motor cortex. The mechanism responsible
for the large width is beyond available theories, which are restricted
to population averages.

Functionally, correlations are important, because they influence the
information contained in the activity of neural populations \citep{Shadlen98,Sompolinsky01a,Morenobote14_1410}.
The recent availability of massively parallel recordings of neural
activity \citep{Stevenson11_139} poses the question whether the joint
statistics allows conclusions on the structure and the operational
regime of the network. For example, a fundamental transition from
regular to chaotic dynamics is known to occur in large networks at
a precisely defined interaction strength, the point at which the regular
state looses linear stability \citep{Sompolinsky88_259}. Highest
computational performance \citep{Crutchfield89_105} is expected at
the edge of this chaotic state \citep{Toyoizumi11_051908}.

Whether and how this critical point is reflected in pairwise correlations
is yet unknown, because the employed mean-field theories \citep{Sompolinsky88_259,molgedey92_3717,Vreeswijk96,Aljadeff15_088101,Kadmon15_041030}
reduce the collective dynamics of the $N$ interacting units to $N$
pairwise independent units each subject to a self-consistently determined
field. Moreover, those theoretical predictions are valid only for
$N\to\infty$. Connections in neuronal networks have, however, limited
range, so that the effective network size is bounded well below the
size of the entire brain. Understanding correlations therefore requires
us to preserve finite-size fluctuations. By a combination of tools
from spin glasses \citep{Crisanti87_4922}, large-$N$ field theory
\citep{Moshe03}, and the functional formalism for classical stochastic
systems by \citet{DeDomincis78_353}, we here obtain a mean-field
theory that reduces the disordered network to a highly symmetric network.
We find that the latter is exposed to external auxiliary fields, whose
fluctuations derive from the quenched disorder of the connections
and explain the neuron to neuron variability.

Employing the formalism, we obtain closed-form expressions for the
mean and width of the distribution of covariances and we explain why
their ratio is $\propto\sqrt{N}$. The dependence on the structural
parameters further allows us to infer network parameters from the
observed activity. The equations establish a link between the distance
to criticality, i.e. the spectral radius of the connectivity matrix,
and the width of the distribution of covariances. The experimental
data introduced in \prettyref{fig:distributions} strongly supports
the operation of the brain close to this critical point.

In the asynchronous irregular regime \citep{Brunel00_183} resembling
cortical activity in the absence of external stimuli, integral covariances
of the fluctuating and correlated network activity $\mathbf{x}(t)$
around the stationary state are given by 

\begin{alignat}{1}
c_{ij} & =\int_{-\infty}^{\infty}\left\langle x_{i}(t+\tau)x_{j}(t)\right\rangle d\tau\nonumber \\
 & =\left[\left(\mathbf{1}-\mathbf{W}\right)^{-1}\mathbf{D}\left(\mathbf{1}-\mathbf{W}^{\mathrm{T}}\right)^{-1}\right]_{ij},\label{eq:covs}
\end{alignat}
following from linear response theory. Here $\mathbf{W}$ is the connectivity
matrix and $\mathbf{D}$ is a diagonal matrix with entries determined
by the stationary mean activities of the neurons. The latter expression
holds for a variety of neuron models \citep{Grytskyy13_131}, such
as binary \citep{Ginzburg94,Dahmen15_arxiv}, leaky integrate-and-fire
\citep{Pernice11_e1002059,Trousdale12_e1002408}, and Poisson model
neurons \citep{Hawkes71_438}. For the spiking models, $c_{ij}$ is
the covariance between spike counts $n_{i}$ and $n_{j}$ (\prettyref{fig:distributions}).
Moreover, \prettyref{eq:covs} is independent of the delays $\mathbf{d}$
and time constants $\boldsymbol{\tau}$ of the system. This correlation
structure can be regarded as originating from the time evolution of
a set of coupled Ornstein-Uhlenbeck processes 
\begin{equation}
\tau_{i}\frac{dx_{i}(t)}{dt}=-x_{i}(t)+\sum_{j=1}^{N}W_{ij}x_{j}(t-d_{ij})+\xi_{i}(t),\label{eq:OUP}
\end{equation}
with uncorrelated zero-mean Gaussian white noise $\boldsymbol{\xi}$
of strength $\mathbf{D}$ \citep{Risken96}.  

Distributions of the covariances $c_{ij}$ \eqref{eq:covs} over different
pairs of neurons therefore arise from distributed stationary mean
activity across neurons (entering $\mathbf{D}$) and from the structural
variability in the disordered connectivity $\mathbf{W}$. We here
focus on the latter and ignore variability in the noise level, which
does not affect the mean integral covariances and only has a minor
effect on the dispersion of integral cross-covariances. 

In nature, the disordered connectivity between neurons is cell-type
specific and distance dependent. However, already a homogeneous random
network \eqref{eq:OUP}, i.e. a network with independent and identically
distributed weights and uniform noise ($\mathbf{D}=D\cdot\mathbf{1}$),
exhibits widely distributed cross-covariances with a small mean (\prettyref{fig:distributions}).
To expose the fundamental mechanism, we here neglect the distance
dependence and cell-type specificity and focus on the effect of the
disordered connectivity alone.

Arbitrary moments of activity variables obeying the Langevin equation
\eqref{eq:OUP} can be derived in the Martin-Siggia-Rose-DeDominicis
path-integral formalism \citep{Martin73,DeDomincis78_353} as functional
derivatives of a generating functional $Z[\mathbf{J}]$ \citep{Chow15}.
In a linear system, all frequencies are independent such that the
generating functional decomposes into a product. The factor for zero
frequency reads 
\begin{equation}
Z[\mathbf{J}]=\int\mD\mathbf{X}\,p(\mathbf{X})\exp\left(\mathbf{J}^{\mathrm{T}}\mathbf{X}\right),\label{eq:def_Z-1}
\end{equation}
where $\int\mD\mathbf{X}=\prod_{j}\int_{-\infty}^{\infty}dX_{j}$
and $p(\mathbf{X})$ is the distribution for the integrated fluctuations
$\mathbf{X}=\mathcal{F}[\mathbf{x}](\omega=0)$. Moments can be obtained
as derivatives with respect to the sources $\mathbf{J}$. $Z[\mathbf{J}]$
in \prettyref{eq:def_Z-1} is given by a Gaussian integral and can
thus be computed analytically 
\begin{eqnarray}
Z[\mathbf{J}] & = & \det\left(\mathbf{1}-\mathbf{W}\right)\int\mD\mathbf{X}\int\mD\tilde{\mathbf{X}}\,e^{S(\mathbf{X},\tilde{\mathbf{X}})+\mathbf{J}^{\mathrm{T}}\mathbf{X}}\label{eq:ZJ}\\
 & = & e^{\frac{1}{2}\mathbf{J}^{\mathrm{T}}\left(\mathbf{1}-\mathbf{W}\right)^{-1}\mathbf{D}\left(\mathbf{1}-\mathbf{W}^{\mathrm{T}}\right)^{-1}\mathbf{J}},\nonumber \\
S(\mathbf{X},\tilde{\mathbf{X}}) & = & \tilde{\mathbf{X}}^{\mathrm{T}}\left(-\mathbf{1}+\mathbf{W}\right)\mathbf{X}+\frac{D}{2}\tilde{\mathbf{X}}^{\mathrm{T}}\tilde{\mathbf{X}},\label{eq:action}
\end{eqnarray}
with response variables $\tilde{\mathbf{X}}$, the measure $\int\mD\tilde{\mathbf{X}}=\prod_{j}\frac{1}{2\pi i}\int_{-i\infty}^{i\infty}d\tilde{X}_{j}$,
and the action $S$.

\selectlanguage{english}%

\selectlanguage{american}%
While covariances between individual neuron pairs depend on the realization
of the random connectivity, we assume that their distribution, in
particular the mean $\overline{c_{ij}}$ and variance $\overline{\delta c_{ij}^{2}}$
across neurons, is self-averaging \citep{Fischer91c}. Exchanging
the order of differentiation and averaging, the disorder averaged
moments $\langle\overline{c_{ij}}\rangle$ can be computed from the
disorder averaged generating function $\left\langle Z[\mathbf{J}]\right\rangle $,
where $\langle\cdot\rangle$ denotes the average across an ensemble
of network realizations with given connectivity statistics. As the
action \eqref{eq:action} for a single realization of $\mathbf{W}$
is quadratic, Wick's theorem applies such that second moments $\langle\overline{c_{ii}^{2}}\rangle=\frac{1}{3}\frac{\partial^{4}}{\partial J_{i}^{4}}\langle Z[J]\rangle$
for any index $i$ and $\langle\overline{c_{ij}^{2}}\rangle=\frac{1}{2}\frac{\partial^{4}}{\partial J_{i}^{2}\partial J_{j}^{2}}\left\langle Z(\mathbf{J})\right\rangle -\frac{1}{2}\langle c_{ii}\rangle^{2}$
for any pair of indices $i\neq j$ can be expressed by fourth derivatives
of $\langle Z(\mathbf{J})\rangle$. 

The generating function formalism allows an algorithmic integration
of the statistics of $\mathbf{W}$. Ignoring insignificant variations
in the normalization of $Z(\mathbf{J})$ \footnote{Variability in the determinant can be accounted for perturbatively,
but only yields subleading contributions which are suppressed for
large network size} and assuming independent and identically distributed entries in $\mathbf{W}$,
the disorder average only affects the coupling term in \prettyref{eq:ZJ}
\begin{eqnarray*}
\left\langle e^{\tilde{\mathbf{X}}^{\mathrm{T}}\mathbf{W}\mathbf{X}}\right\rangle  & = & \prod_{i,j}\phi(\tilde{X}_{i}X_{j})=\prod_{i,j}\exp\left(\sum_{k=1}^{\infty}\frac{\kappa_{k}}{k!}(\tilde{X}_{i}X_{j})^{k}\right).
\end{eqnarray*}
In the resulting cumulant expansion \citep{DeDomincis78_353,Nishimori01_01}
$\phi$ is the characteristic function for a single connection $W_{ij}$
and $\kappa_{k}$ its $k$-th cumulant \citep{Gardiner85}. Independence
of network size can only be expected, if the fluctuations of the input
to a neuron are independent of $N$, requiring synaptic weights to
scale with $1/\sqrt{N}$ \citep{Vreeswijk96,VanVreeswijk98_1321},
such that the cumulant expansion is an expansion in $1/\sqrt{N}$.
In an Erd\H{o}s-R\'enyi network, a single connection is drawn from
a Bernoulli distribution $\mathcal{B}(p,w)$ with connection probability
$p$ and weight $w=N^{-\frac{1}{2}}w_{0}$. A truncation at the second
cumulant ($\propto N^{-1}$) maps $\mathbf{W}$ to a Gaussian connectivity
$\mathcal{N}(\mu,\sigma^{2}/N)$ with $\mu_{ij}=\mu=pw_{0}N^{-\frac{1}{2}}$
and $\sigma^{2}=p(1-p)w_{0}^{2}$ so that 
\begin{eqnarray*}
\left\langle Z[\mathbf{J}]\right\rangle  & \sim & \int D\mathbf{X}\int D\tilde{\mathbf{X}}\,e^{S_{0}(\mathbf{X},\tilde{\mathbf{X}})+S_{\mathrm{int}}(\mathbf{X},\tilde{\mathbf{X}})+\mathbf{J}^{\mathrm{T}}\mathbf{X}},\\
S_{0}(\mathbf{X},\tilde{\mathbf{X}}) & = & \tilde{\mathbf{X}}^{\mathrm{T}}\,\left(-\mathbf{1}+\boldsymbol{\mu}\right)\,\mathbf{X}+\frac{D}{2}\tilde{\mathbf{X}}^{\mathrm{T}}\tilde{\mathbf{X}},\\
S_{\mathrm{int}}(\mathbf{X},\tilde{\mathbf{X}}) & = & \frac{\sigma^{2}}{2N}\tilde{\mathbf{X}}^{\mathrm{T}}\tilde{\mathbf{X}}\,\mathbf{X}^{\mathrm{T}}\mathbf{X}.
\end{eqnarray*}
The second cumulant ($\sigma^{2}/N$) is the first non-trivial contribution
to the second moment of covariances. While higher cumulants of the
connectivity have an impact on higher moments of the distribution
of covariances, their effect on the first two moments is suppressed
by the large network size. 

\begin{figure}
\centering{}\includegraphics[scale=0.4]{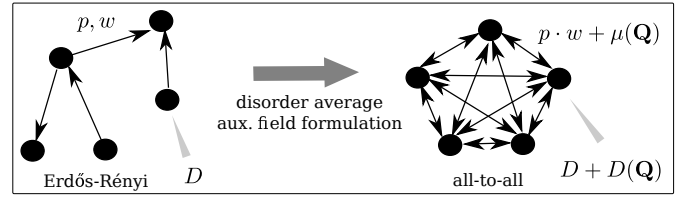}\caption{Disorder average maps network with frozen variability in connections
to highly symmetric network on the background of fluctuating auxiliary
fields $\mathbf{Q}$, which induce additional temporal variability
(noise $D(\mathbf{Q})$) and global variability in connections ($\mu(\mathbf{Q})$).
\label{fig:mapping}}
\end{figure}

The interaction term $S_{\mathrm{int}}$ prevents an exact calculation
of the disorder-averaged generating function. A converging perturbation
series can be obtained in the auxiliary field formulation \citep{Moshe03},
where a field $Q_{1}=\frac{\sigma^{2}}{N}\mathbf{X}^{\mathrm{T}}\mathbf{X}$
is introduced for the sum of a large number of statistically equivalent
activity variables. Using the Hubbard-Stratonovich transformation
\begin{eqnarray*}
e^{S_{\mathrm{int}}(\mathbf{X},\tilde{\mathbf{X}})} & \sim & \int D\mathbf{Q}\;e^{-\frac{N}{\sigma^{2}}Q_{1}Q_{2}+\frac{1}{2}Q_{1}\tilde{\mathbf{X}}^{\mathrm{T}}\tilde{\mathbf{X}}+Q_{2}\mathbf{X}^{\mathrm{T}}\mathbf{X}},
\end{eqnarray*}
one obtains a free theory, i.e. quadratic action in the activity ($\mathbf{X}$)
and response variables ($\tilde{\mathbf{X}})$, on the background
of fluctuating fields $\mathbf{Q}$
\begin{eqnarray}
\left\langle Z[\mathbf{J}]\right\rangle  & \sim & \int D\mathbf{Q}\,e^{-\frac{N}{\sigma^{2}}Q_{1}Q_{2}+\ln\left(Z_{\mathbf{Q}}[\mathbf{J}]\right)},\label{eq:mean_Z}\\
Z_{\mathbf{Q}}[\mathbf{J}] & = & \int D\mathbf{X}\int D\tilde{\mathbf{X}}\;e^{S_{0}(\mathbf{X},\tilde{\mathbf{X}})+\frac{1}{2}Q_{1}\tilde{\mathbf{X}}^{\mathrm{T}}\tilde{\mathbf{X}}+Q_{2}\mathbf{X}^{\mathrm{T}}\mathbf{X}+\mathbf{J}^{\mathrm{T}}\mathbf{X}}.\nonumber 
\end{eqnarray}
The large dimensional integrals of the free theory $Z_{\mathbf{Q}}[\mathbf{J}]$
can be solved analytically yielding a two-dimensional interacting
theory in the auxiliary fields $Q_{1}$ and $Q_{2}$. The auxiliary
field formalism translates the high-dimensional ensemble average over
$\mathbf{W}$ to a low-dimensional average over $\mathbf{Q}$, i.e.
a mapping of the local disorder in the connections to fluctuations
of the global connection strength and noise level in a highly symmetric
all-to-all connected network, illustrated in \prettyref{fig:mapping}
\footnote{Note that this follows after integrating over the variables $\tilde{\mathbf{X}}$.
The result is equivalent to a noise strength $D(\mathbf{Q})$ and
a connection strength $\mu(\mathbf{Q})$ that depend on $Q_{1}$ and
$Q_{2}$.}. Only in the special case of vanishing mean connection strength $\mu=0$,
the system factorizes into $N$ unconnected units, each interacting
with the same set of fields $\mathbf{Q}$. The all-to-all network
not only captures the auto-covariance of a single neuron, but also
the cross-covariance with any other neuron.

A loopwise expansion \citep{Sompolinsky81,Sompolinsky82_6860,ZinnJustin96}
of the exponent in \prettyref{eq:mean_Z} around self-consistently
determined and source-dependent saddle points $\mathbf{Q}_{\mathbf{J}}^{*}$
of the $\mathbf{Q}$-integrals yields a $1/N$ expansion of $\left\langle Z(\mathbf{J})\right\rangle $.
For large networks, the zeroth order (tree-level in $\mathbf{Q}$)
is sufficient to calculate the two- and four-point correlators that
yield the leading order contributions to the mean integral covariances
\begin{eqnarray}
\overline{c_{ij}} & = & \left[\left(\mathbf{1}-\boldsymbol{\mu}\right)^{-1}\frac{D}{1-R^{2}}\left(\mathbf{1}-\boldsymbol{\mu}^{\mathrm{T}}\right)^{-1}\right]_{ij}=D_{\mathrm{r}}\gamma_{ij}\label{eq:mean_cij}
\end{eqnarray}
and the variance of integral covariances
\begin{eqnarray}
\overline{\delta c_{ij}^{2}} & = & R^{2}\left[\frac{1}{\left(1-R^{2}\right)^{2}}+\frac{1}{1-R^{2}}\right]D_{\mbox{\ensuremath{\mathrm{r}}}}^{2}\chi_{ij},\label{eq:var_cij}
\end{eqnarray}
with $\gamma_{ij}=\delta_{ij}+\gamma$, $\chi_{ij}=\frac{1}{N}\left(1+\delta_{ij}+\mathcal{O}(1/N)\right)$,
$\gamma=\mathcal{O}(1/N)$, which depend on the deterministic network
structure, the spectral radius $R=\sqrt{1+\gamma}\,\sigma$ of the
connectivity matrix $\mathbf{W}$ \citep{Rajan06}, and the noise
strength $D_{\mathrm{r}}=D+D(\mathbf{Q}_{0}^{\ast})$. The latter
is renormalized by the structural variability ($D(\mathbf{Q}_{0}^{\ast})=D\frac{R^{2}}{1-R^{2}}$,
see \prettyref{fig:mapping}). The average connection strength, however,
is not affected ($\mu(\mathbf{Q}_{0}^{\ast})=0$, see \prettyref{fig:mapping}).
While the previous expressions are obtained from saddle points evaluated
at vanishing sources $\mathbf{Q}_{0}^{\ast}=\mathbf{Q}_{\mathbf{J}=0}^{\ast}$,
the distinct dependence of $D(\mathbf{Q}_{\mathbf{J}}^{\ast})$ and
$\mu(\mathbf{Q}_{\mathbf{J}}^{\ast})$ on the external sources $\mathbf{J}$
reflects fluctuations of auxiliary fields $\mathbf{Q}$ across network
realizations. These fluctuations in $Q_{1}$ and $Q_{2}$ contribute
each one term to the dispersion of covariances in \prettyref{eq:var_cij}
with different scaling in $R$. This source dependence has been neglected
in the pioneering work introducing the functional formulation of disordered
systems \citep{Sompolinsky82_6860}.

The mean connection strength $\mu$ determines $\gamma$ and $\chi_{ij}$
and acts as a negative feedback in inhibitory or inhibition-dominated
networks \citep{Tetzlaff12_e1002596}. While this feedback suppresses
mean cross-covariances \eqref{eq:mean_cij}, it only yields a subleading
contribution to the dispersion \eqref{eq:var_cij}. The spread of
individual cross-covariances is therefore predominantly determined
by fluctuations in connection weights. These fluctuations cause broad
distributions of cross-covariances of both signs even in a homogeneous
network \footnote{A decomposition of fluctuations into the population-averaged fluctuation
and orthogonal modes yields that the population-averaged fluctuation,
which exclusively determines average covariances, is suppressed by
the average inhibitory connection weight $\mu<0$, while the second
moment of covariances results from the fluctuations of all remaining
modes, which are not suppressed by negative feedback.}. 

\begin{figure}
\raggedright{}\includegraphics{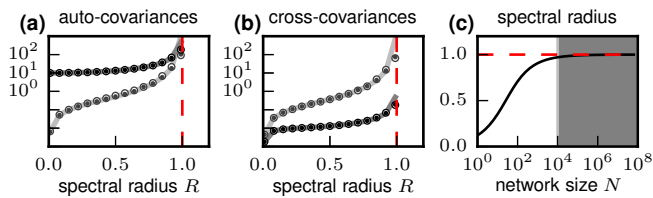}\caption{Mean (dark gray) and standard deviation (light gray) of integral auto-
(a) and cross-covariances (b) for different spectral radii $R$. Solid
curves indicate analytical predictions, symbols show numerical results
for one realization at each parameter setting (dots: Gaussian connectivity;
open circles: Erd\H{o}s-R\'enyi connectivity; network size $N=1000$).
(c) Predicted spectral radius of the effective connectivity of the
macaque motor cortex network as a function of network size for given
moments of experimentally observed parameters of the covariance distribution
(\prettyref{fig:distributions}, $\overline{c_{ij}}=0.12$, $\overline{\delta c_{ij}^{2}}=7.89$,
$\overline{c_{ii}}=16.16$, $\overline{\delta c_{ii}^{2}}=432.89$).
The shaded area marks the range of biologically plausible effective
network sizes corresponding to the spatial scale of the recordings.
The red dashed line indicates the critical point $R=1$.\label{fig:theory_check}}
\end{figure}

At a critical point $R=1$ (\prettyref{fig:theory_check}), where
the linear system becomes unstable due to the largest eigenvalue of
the connectivity matrix $\mathbf{W}$ exceeding unity, we observe
a divergence of auto- and cross-covariances. In the slightly sub-critical
regime, slowly decaying fluctuations generate individual covariances
in the network much larger than the average across neurons. The same
correlation structure exists in a nonlinear network \citep{Sompolinsky88_259}
with infinitesimal additive noise, leading to fluctuating activity
also in the regular regime. In this model, the point of linear instability
coincides with a transition to chaos and a divergence in topological
complexity \citep{wainrib13_118101}. 

The analytic expressions for the moments of covariances Eqs. \eqref{eq:mean_cij}
and \eqref{eq:var_cij} can be used to infer network parameters from
experimentally observed covariance distributions. For the highly symmetric
network considered here, the application of Wick's theorem yields
at leading order a trivial factor two between the variance of integral
auto- and cross-covariances (see definition $\chi_{ij}$ above) and
thus requires one free parameter in the inversion of Eqs. \eqref{eq:mean_cij}
and \eqref{eq:var_cij}. For given network size $N$, the spectral
radius of the effective connectivity $\mathbf{W}$ is predominantly
determined by the width of the distribution of cross-variances normalized
by the mean auto-covariances

\begin{eqnarray}
R^{2} & \approx & 1-\sqrt{\frac{1}{1+N\frac{\overline{\delta c_{ij}^{2}}}{\overline{c_{ii}}^{2}}}}.\label{eq:radius}
\end{eqnarray}
Distributions of covariances can therefore be used to infer the operational
regime of the network, i.e. the distance to criticality. \prettyref{fig:theory_check}c
shows that for the distribution of covariances measured in macaque
motor cortex (\prettyref{fig:distributions}) and biologically plausible
network sizes, the linear network model must operate close to criticality
to explain the data. The small change of the spectral radius in the
biologically relevant range further illustrates the robustness of
the result with respect to a potential bias of the experimental estimates
due to limited observation time or the number of recorded neurons
\citep{Ecker10}. 

Mean integral auto-covariances \eqref{eq:mean_cij} and the variance
of integral cross-covariances \eqref{eq:var_cij} are predominantly
determined by the global noise level and the spectral radius. Therefore
these measures are rather insensitive to specific deterministic features
of the connectivity and distributions of noise amplitudes across neurons.
This suggests that \prettyref{eq:radius} also holds qualitatively
for more complicated network topologies and variable noise levels. 

In a biologically plausible Erd\H{o}s-R\'enyi network, randomness
in connections is controlled by the weight of non-zero connections.
Tuning these weights of the effective connectivity by, for example,
plasticity mechanisms or external inputs to the network, one can adjust
the overall correlation structure and drive the network into a linearly
unstable regime with large transients introduced by external perturbations.

One can use Eqs. \eqref{eq:mean_cij} and \eqref{eq:var_cij} to uniquely
determine the parameters of a homogeneous network of arbitrary size
that generates distributions of covariances matching the first two
moments of experimental data (\prettyref{fig:distributions}). An
exception is the variance of integral auto-covariances that is sensitive
to the inter-neuron variability of the noise level. A better agreement
between the higher-order moments of the experimental distributions
for auto- and cross-covariances and the model results requires more
realistic network topologies, such as excitatory and inhibitory populations
driven by heterogeneous noise and spatially dependent connectivity.
We have high hopes that generalizations of the formalism will turn
out to be straight forward.

\paragraph{}

\begin{acknowledgments}
We are grateful to Alexa Riehle and Thomas Brochier for providing
the experimental data and to Sonja Grün for fruitful discussions on
their interpretation. We thank Vahid Rostami for helping us with the
analysis of the multi-channel data. The research was carried out within
the scope of the International Associated Laboratory ``Vision for
Action - LIA V4A`` of INT (CNRS, AMU), Marseilles and INM-6, Jülich.
This work was partially supported by HGF young investigator's group
VH-NG-1028, Helmholtz portfolio theme SMHB, and EU Grant 604102 (Human
Brain Project, HBP). 
\end{acknowledgments}

\end{document}